\UseRawInputEncoding

\documentclass[10pt,conference]{IEEEtran}
\IEEEoverridecommandlockouts
\usepackage{amsmath,amssymb,amsfonts}
\usepackage{algorithmic}
\usepackage[linesnumbered, ruled]{algorithm2e}
\usepackage{graphicx}
\usepackage{textcomp}
\usepackage{multirow}
\usepackage{diagbox}
\usepackage[table,xcdraw]{xcolor}
\usepackage{array}
\usepackage{hhline}
\usepackage{svg}

\newcolumntype{L}[1]{>{\raggedright\arraybackslash}p{#1}}

\def\BibTeX{{\rm B\kern-.05em{\sc i\kern-.025em b}\kern-.08em
    T\kern-.1667em\lower.7ex\hbox{E}\kern-.125emX}}
\begin{document}

\title{
Multidomain transformer-based deep learning for early detection of network intrusion
}

\author{
\IEEEauthorblockN{Jinxin Liu$^1$, Murat Simsek$^1$, Michele Nogueira$^2$, Burak Kantarci$^1$}
\IEEEauthorblockA{\textit{$^1$School of Electrical Engineering and Computer Science} \\
\textit{University of Ottawa, Ottawa, ON, Canada}\\
\textit{$^2$Department of Computer Science} \\
\textit{Universidade Federal de Minas Gerais, MG, Brazil}\\
$^1$\{jliu367, murat.simsek, burak.kantarci\}@uottawa.ca,~$^2$michele@dcc.ufmg.br}
}

\maketitle

\begin{abstract}
Timely response of Network Intrusion Detection Systems (NIDS) is constrained by the flow generation process which requires accumulation of network packets.
This paper introduces Multivariate Time Series (MTS) early detection into NIDS 
to identify malicious flows prior to their arrival at target systems. With this in mind, we first propose a novel feature extractor, Time Series Network Flow Meter (TS-NFM), that represents network flow as MTS with explainable features, and a new benchmark dataset is created using TS-NFM and the meta-data of CICIDS2017, called SCVIC-TS-2022. Additionally, a new deep learning-based early detection model called Multi-Domain Transformer (MDT) is proposed, which incorporates the frequency domain into Transformer. This work further proposes a Multi-Domain Multi-Head Attention (MD-MHA) mechanism to improve the ability of MDT to extract better features. Based on the experimental results, the proposed methodology improves the earliness of the conventional NIDS (i.e., percentage of packets that are used for classification) by $5\times10^4$ times
and duration-based earliness (i.e., percentage of duration of the classified packets of a flow) by a factor of 60, resulting in a 84.1\% macro F1 score (31\% higher than Transformer) on SCVIC-TS-2022. Additionally, the proposed MDT outperforms 
the state-of-the-art early detection methods by 5\% and 6\% on ECG and Wafer datasets, respectively.
\end{abstract}


\section{Introduction}
A Network Intrusion Detection System (NIDS) is 
a critical mechanism that protect networks against a variety of threats by inspecting network traffic \cite{10.1145/3530812}. A NIDS typically builds on a three-step workflow: collecting network packets to construct flows, extracting features, and making decisions \cite{9653662}.
Numerous machine/deep learning models have been proposed, achieving remarkable detection performance with the rapid response. However, the process of flow formation and feature extraction requires relatively long time to accumulate distinguishable features; for example, flows are collected within a 2-minute and 1 minute time window on CICIDS2017~\cite{sharafaldin_toward_2018} and UNSW-NB15~\cite{moustafa_unsw-nb15_2015}, respectively. As a result, a substantial amount of malicious traffic traces (particularly Advanced Persistent Threat (APT) traffic) are detected after it reaches target systems~\cite{10.1145/3530812}.
APT attacks or penetration tests typically involve the execution of malicious payloads, which is one of the crucial steps of an APT attack.

Table~\ref{tab:payloads} summarizes the number of packets and time required to execute 
common payloads. Due to the powerful functionalities of meterpreter payloads (a well-known payload for penetration tests), they need more packets and time to establish communication tunnels, and shell/PowerShell payloads require only 10 packets in 0.1s to execute. However, in conventional NIDS configurations, even if a meterpreter is utilized, attackers have sufficient time to conduct post-infiltration activities (such as data collection or lateral movement) before being discovered. Early detection in NIDS, which aims to detect intrusions as early as possible using the initial subsequences of a time series, could be ideal to mitigate the impact of cyber-physical attacks. One of the essential criteria in this study is to determine the length of the subsequence over the length of the time series for classification. In the context of NIDS, earliness is the ratio of the number of packets used for intrusion detection to the total number of packets in a flow. As seen in Table~\ref{tab:payloads}, the duration of network intrusion executions is also crucial. Consequently, we propose a duration-based earliness, which is defined as the duration of packets throughout the duration of a flow.

\begin{figure}
    \centering
    \includegraphics[width=.8\linewidth,trim = 0.0cm 0.2cm 0.0cm 0.0cm, clip]{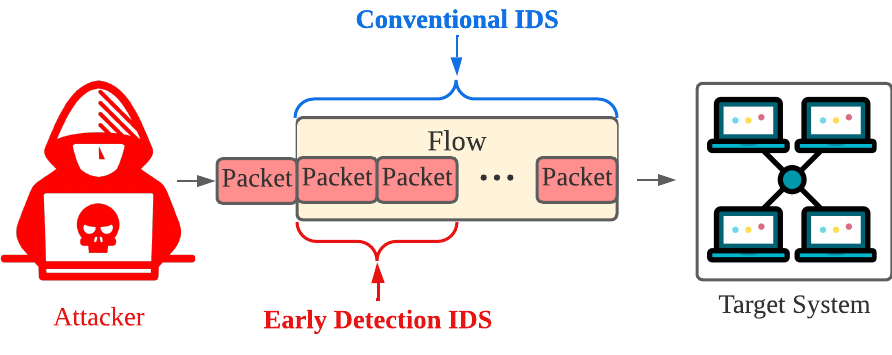}
    \caption{
    An Illustrative Comparison of a Conventional NIDS and an Early Detection IDS 
    }
    \label{fig:compared with coventional nids}
\end{figure}

This work introduces early detection into the NIDS field, allowing NIDSs to detect intrusions before they completely reach target systems, placing NIDSs one step ahead of attacks. 
This study proposes a novel framework to represent network flows as multivariate time series named Time Series Network Flow Meter (TS-NFM). Additionally, a new early detection model, Multi-Domain Transformer (MDT), is proposed 
to significantly improve NIDSs' earliness without losing the detection performance. 
Fig.~\ref{fig:compared with coventional nids} shows a conventional NIDS detecting intrusions after the entire flow has been formed, whereas our proposed Network Intrusion Early Detection System is capable of detecting intrusions with only the first few packets, providing organizations with sufficient time to respond and safeguard the target systems by cutting off malicious traffic before their arrival and the occurrence of any damage. This paper differs from the state of the art in NIDS with four key contributions as listed below:
\begin{itemize}
    \item To the best of our knowledge, this is the first study that leverages multivariate time series-based early detection for a NIDS.
    \item A novel network intrusion feature extractor called Time Series Network Flow Meter (TS-NFM) is proposed to extract features from network flows and present them in multivariate time series format.
    This work further forms a new dataset (SCVIC-TS-2022)\footnote{Available at IEEE Dataport: https://dx.doi.org/10.21227/qm9h-8c05} by utilizing the meta-data (i.e., network packets and labelling information) from a well known benchmark dataset CICIDS2017.
    
    \item We propose the Multi-Domain Transformer (MDT), which outperforms baseline models and improves the Transformer \cite{vaswani_attention_2017} classifier's performance on not only SCVIC-TS-2022 but also other early detection datasets.
    \item This work proposes Multi-Domain Multi-Head Attention (MD-MHA) mechanism which can support the MDT in the extraction of useful features from MTS data.
\end{itemize}


This paper proceeds as follows. 
Section~\ref{sec:related work} summarizes the previous work related to NIDSs and early detection. Section~\ref{sec:proposed method} details the proposed methodology, which includes TS-NFM and MDT. Section~\ref{sec:experiments} discusses the datasets and experimental results of MDT. Finally, Section~\ref{sec:conclusion} draws the conclusions regarding the contributions of this study.

\begin{table}[]
\centering
\caption{Transmission Duration and the Number of Packets of Different Types of Payloads. The Target Machine (Windows Server 2012 R2) is Exploited via PsExec from Metasploit. (PSExec needs extra 88 packets for 2.13s)}
\label{tab:payloads}
\resizebox{\linewidth}{!}{%
\begin{tabular}{|l|l|l|l|}
\hline
Payload Type                 & Connection Type & Num of packets & Duration (s) \\ \hline
\multirow{4}{*}{meterpreter} & bind\_tcp       & 437            & 2.33         \\ \cline{2-4} 
                             & reverse\_tcp    & 454            & 2.34         \\ \cline{2-4} 
                             & reverse\_http   & 267            & 31.80        \\ \cline{2-4} 
                             & reverse\_https  & 563            & 31.41        \\ \hline
\multirow{2}{*}{shell}       & bind\_tcp       & 12             & 0.78         \\ \cline{2-4} 
                             & reverse\_tcp    & 10             & 0.10         \\ \hline
\multirow{2}{*}{powershell}  & bind\_tcp       & 11             & 1.70         \\ \cline{2-4} 
                             & reverse\_tcp    & 10             & 0.10         \\ \hline
\end{tabular}%
}
\end{table}

\section{Related Work}
\label{sec:related work}


\subsection{Machine Learning-based NIDSs}
Resende et al.~\cite{resende_adaptive_2018} proposed an anomaly-based intrusion detection system that is optimized using a genetic algorithm. The work further integrates supervised and unsupervised approaches to achieve high performance under the CICIDS2017 dataset.
%

Marir et al.~\cite{marir_distributed_2018} proposed a distributed intrusion detection system that combines Deep Belief Networks (DBN) and Support Vector Machines (SVM) to detect attacks in large volumes of traffic. The proposed strategy is able to significantly increase the baseline performance.
Min et al.~\cite{min_su-ids_2018} proposed the SU-IDS AutoEncoder network, which learns from unlabeled examples using semi-supervised and unsupervised methods. Xu et al.~\cite{xu_method_2020} introduced few-shot learning into NIDSs. By extracting image-like features from raw network packets, and feeding these features into FC-Net allows NIDS to detect new types of intrusions with limited samples without significant compromise in the detection performance. Marteau et al.~\cite{marteau_random_2021} proposed DiFF-RF, a semi-supervised technique for detecting anomalies in network flows that overcomes the constraints of Isolation Forest. DiFF-RF is tested against a variety of network intrusion datasets.
Chen et al. \cite{9685318} proposed an ensemble approach named All Predict Wisest Decides (APWD) that trains various machine learning models and relies on the decisions of specific ML models only if they outperform the rest of the classifiers for certain traffic types. 
Kim et al.~\cite{kim_robust_2022} proposed an early detection framework based on statistical machine learning. However, their proposed framework not only requires the arbitrary setting of the number of packets used for detection for each class, but also cannot handle minor classes, i.e., classes with a small number of samples.

\subsection{Early Detection on Time Series Data}

Ding et al. \cite{ding_querying_2008} employ  1-Nearest Neighbor Classification (1NN) to multi-variate time series data. However, extracting useful features from time series remains a challenge. Hence, shorter inference times are desired.
Ghalwash et al.~\cite{ghalwash_early} proposed a method named Multivariate Shapelets Detection (MSD), which is a shapelet-based algorithm that incorporates HMM and SVM models. The proposed MSD is capable of identifying critical decision-making segments. Lin et al.~\cite{lin_reliable_2015} proposed the REACT methodology for effectively classifying multivariate time series with a variety of features on multi-GPU hardware settings.
Huang et al.~\cite{huang_multivariate_2018} propose the Multi-Domain DNN (MDDNN), which analyzes time series in both the time and frequency domains and incorporates CNN-LSTM to extract feature representations.
Hsu et al.~\cite{hsu_multivariate_2019} proposed an 
Explainable Time. Series Classification Model (ETSCM)
based on MDDNN that utilizes an attention mechanism 
to make decisions based on segments with high confidence score. ETSCM is capable of identifying critical components of time series, hence increasing the model's interpretability.

Previous research in both domains (i.e., NIDS and early detection) has yielded promising results; however, they have not been integrated to improve the earliness of a NIDS. Thus, this paper aims to bridge the gap between early detection and NIDSs so as to enable an NIDS to respond quickly before the entire flow is formed so that attacks are intercepted before they reach target systems.

\section{Network Intrusion Early Detection System}
\label{sec:proposed method}

This section presents and elaborates on the proposed method for the early detection of network intrusions. To begin with, a new network feature extractor is presented to characterize a network flow in a time-series format in order to identify them prior to their complete arrival. Additionally, inspired by MDDNN, which demonstrates that frequency domain can improve the performance of deep learning models, we propose a Multi-Domain Transformer (MDT) to detect network intrusions using the early packets of flows. MDT integrates Fourier transformation with a transformer encoder that learns additional information from the frequency domain.

\subsection{Problem Definition}

Traditionally, a machine learning-based NIDS detects network intrusions by collecting network packets to form flows, extracting features from the flows to generate a tabular dataset, and feeding the flow-based tabular dataset to machine learning algorithms, whereas this paper is concerned with detecting intrusions as early as possible. Two important issues should be addressed to accomplish this goal: 1) representing flows as time series, 2) detecting intrusions using only the beginning subsequence of time series.

\begin{figure}
    \centering
    \includegraphics[width=.8\columnwidth, trim = 0.0cm 0.3cm 0.0cm 0.0cm, clip ]{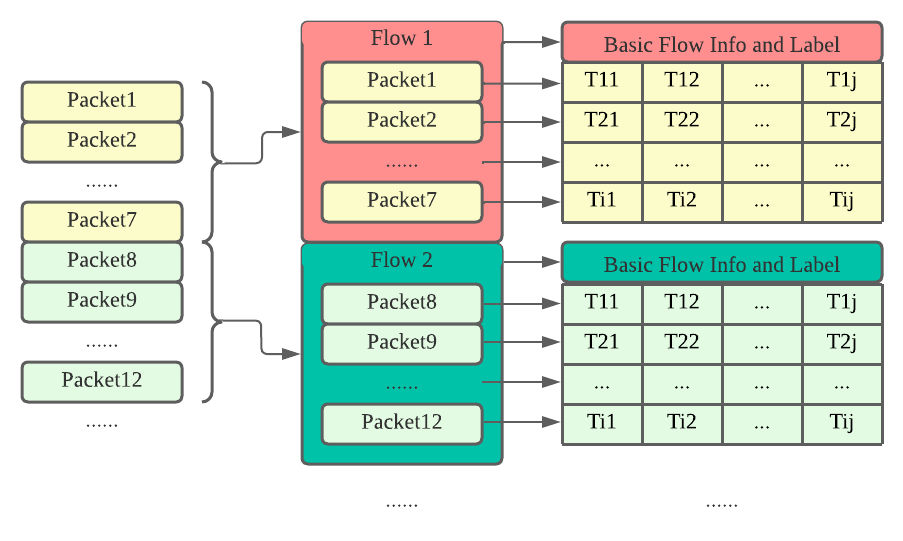}
    \caption{
    Time Series Representation of Network Flow Meter Features
    }
    \label{fig:features}
\end{figure}

Conventional flow-based feature extractors can successfully obtain features from flows but they encounter earliness limitations as features are observed through a completed time window. Hence, we propose extracting packet-based features from the network flows forming time-series datasets.
Each flow $f$ is defined as a set of packets $P=\{p_1, p_2, ..., p_L\}$ with the same session key: source IP, source port, destination IP, destination port, and time window.
From each flow, TS-NFM extracts a Multivariate Time Series (MTS) $\mathbb{R}^{L*d}$ which consists of features from packets, where $L$ is the total length of the MTS and $d$ is the number of features extracted from a packet. To classify intrusions in early stages, only the Early Subsequence (ES) $\mathbb{R}^{l*d}$ is used for classification, where $l < L$. 

Besides classification performance, another important metric of early detection is the Earliness, which is formulated as $E=\frac{l}{L}$. As shown in Table \ref{tab:payloads}, some payloads or attacks may have less number of packets while being executed for longer duration than others, such as Reverse TCP and Reverse HTTP in Meterpreter. Therefore, besides earliness, a duration-based earliness is considered to prevent intrusions with low number of packets but long duration of launching period.
Duration-based Earliness (DE) can be defined as $DE=\frac{Dur(ES)}{Dur(MTS)}$. Since E and DE should be considered together, we further propose Flow Earliness ($FE=(E, DE)$) as a comprehensive metric for network intrusion early detection.

\subsection{Time Series Network Flow Meter}

A Conventional NIDS feature extractor (NIDSFE), such as CICFlowMeter \cite{habibi_lashkari_characterization_2017} or NFStream \cite{aouini_nfstream_2022}, extracts features as a combination of the three aspects (i.e., direction, feature kind and statistical characteristics) 
. For example, Forward + Packet Bytes + Min specifies the flow's minimum packet bytes. The primary constraint on the earliness of the classical NIDSFE is not only the extraction of  features after the generation of the flow but also the fact that those basic statistical characteristics require a certain number of packets to adequately describe a flow. Thus, rather than manually summarizing flow properties using statistical characteristics, we propose extracting features directly from packets constituting an MTS and allowing machine learning models to learn, define and classify flows.

To extract MTS from a network flow, we propose a time series-based feature extractor named Time Series Network Flow Meter (TS-NFM). As shown in Fig. \ref{fig:features}, packets with the same session key (i.e., source IP, source port, destination IP, destination port, time window) are grouped together to form a flow. Instead of extracting statistical features from a flow, TS-NFM extracts features from each packet inside a flow. The packet-based features include direction, Inter-Arrival Time (IAT), number of bytes, and TCP flags. In order to label network flows, the flow session information (basic info) is stored in tabular format whereas the features are stored in MTS format, i.e., $\mathbb{R}^{L*d}$, where $L$ is the number of packets inside a flow, and $d$ is the number of features extracted from each packet.



\subsection{Multi-Domain Transformer}

    

\begin{figure}
    \centering
    \includegraphics[width=\columnwidth]{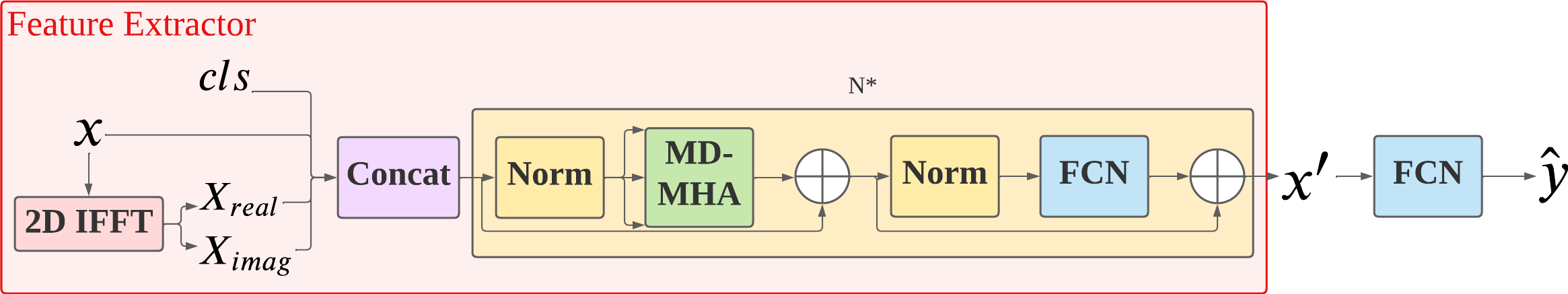}
    \caption{MDTransformer Architecture}
    \label{fig:mdt}
\end{figure}

\begin{figure}
    \centering
    \includegraphics[width=\linewidth]{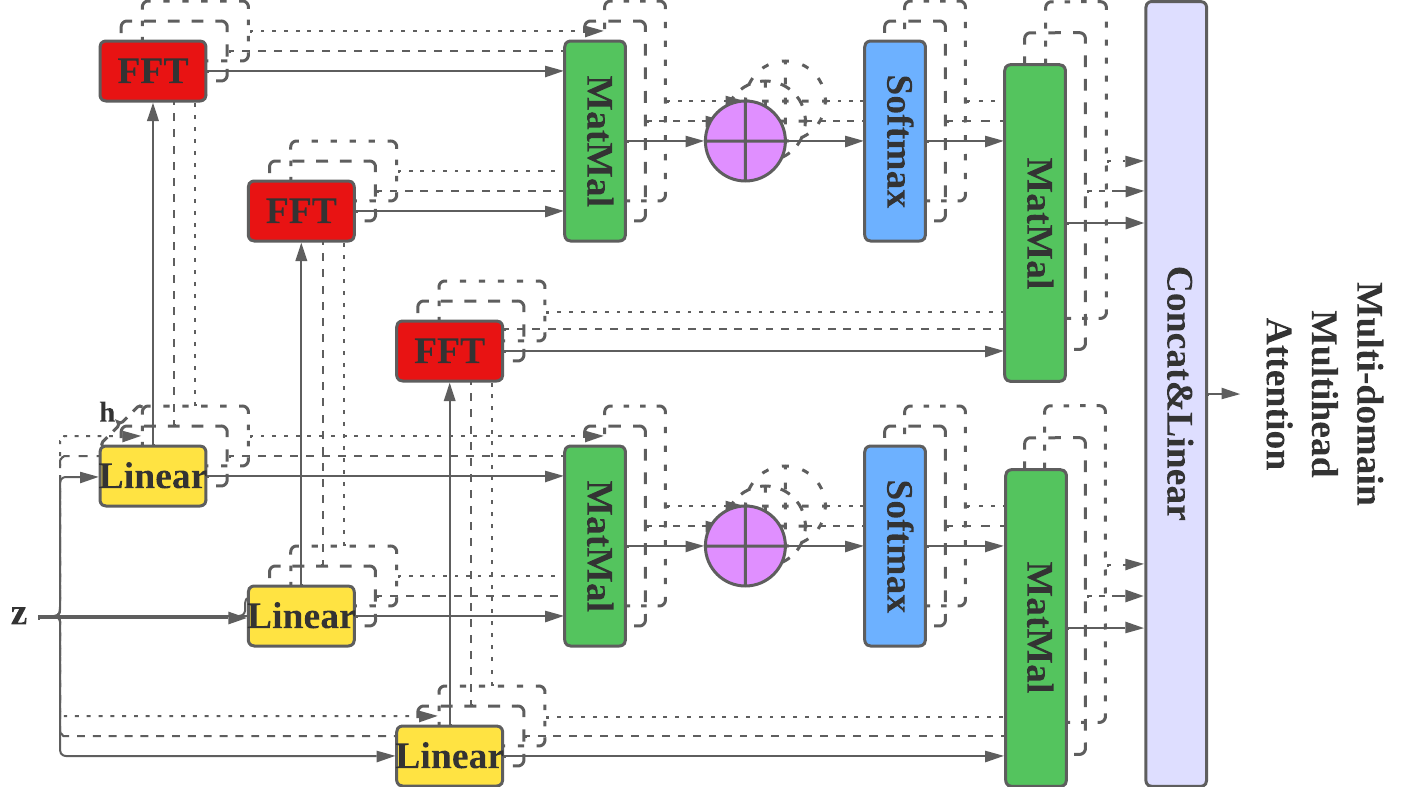}
    \caption{
    Multi-Domain Multi-Head Attention (MD-MHA). Comparing with the vanilla attention, we obtain the attention score from both time domain and frequency domain to obtain more information.
    }
    
    \label{fig:mdmha}
\end{figure}

\begin{figure}
    \centering
    \includegraphics[width=\linewidth]{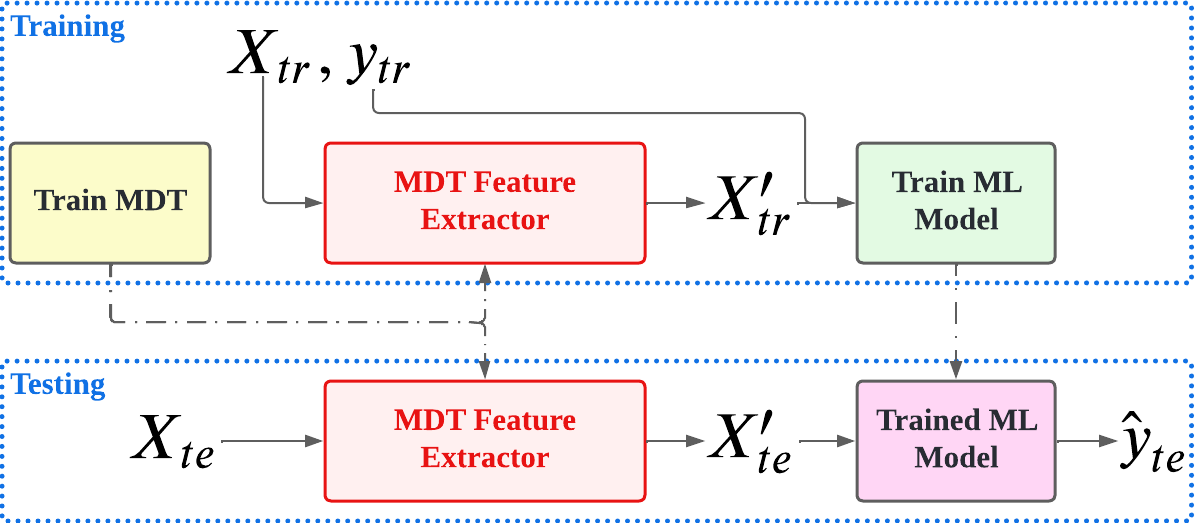}
    \caption{MDT Using ML Classifier}
    \label{fig:ml classifier}
\end{figure}


This section introduces the Multi-Domain Transformer (MDT), a deep learning model for early classification that incorporates the frequency domain into the original Transformer model. 

Each input of the early classification problem contains limited information; hence, it is essential to extract useful features from the MTS. The frequency domain 
provides extra information alongside the time domain of time series. The architecture of MTD is depicted in Fig.~\ref{fig:mdt}. 

An MTS is initially converted using two-dimensional  Inverse Fast Fourier Transform (IFFT). The outputs of the IFFT (i.e., the real and imaginary sections) are concatenated with MTS and fed into a Transformer encoder \cite{devlin_bert_2019}.
IFFT is required for the following reason. For MTS with a small number of dimensions/features $d$, the layer normalization of the Transformer encoder makes time series difficult to differentiate. For example, the ECG dataset contains only two dimensions. After layer normalization, all time series are changed to [0, 0], [1, -1], and [-1, 1], which means that all amplitude information in MTS is lost. As a result, the dimensions can be increased by using IFFT. MDDNN makes use of the one-dimensional IFFT; however, the imaginary part of the one-dimensional IFFT may be zero. Thus,  a 2D IFFT is used with the real and imaginary components concatenated with the original input in order to further increase the number of dimensions. Additionally, 2D IFFT may extract a greater amount of local information from MTS than 1D IFFT.

This work proposes a Multi-Domain Multi-Head Attention (MD-MHA) approach for further enhancements on the transformer encoder. Fig. \ref{fig:mdmha} depicts the architecture of MD-MHA where $z$ is used as an input of the MD-MHA. Three linear layers ($W_Q$, $W_K$, $W_V$) are fed into $z$ to obtain $Q$, $K$, $V$ that are further converted to the frequency domain $Q'=FFT(Q)$, $K'=FFT(K)$, $V'=FFT(V)$. MD-MHA follows 
as in Eq.~\ref{eq:mdmha}.
\begin{equation}
\label{eq:mdmha}
    MD-MHA(z) = [h_1. h_2, ..., h_h, h'_1, h'_2, ... h'_h]W_o,
\end{equation}
 where
\begin{equation*}
    h_i = softmax(\frac{QK^T}{\sqrt{d_{model}}})V \text{, } h'_i= softmax(\frac{Q'K'^T}{\sqrt{d_{model}}})V'
\end{equation*}

As FFT transforms the time domain to the frequency domain, it can explicitly provide MDT with a variety of views. MD-MHA increases the number of heads without increasing the number of parameters, hence decreasing the possibility of overfitting. 

Network intrusion detection datasets contains imbalanced classes; hence, machine learning models (such as XGBoost) perform better than deep learning-based classifiers including Deep MLP and TabNet on tabular datasets \cite{liu_collaborative_2022}. Therefore, this work incorporates machine learning models into MDT. As Fig. \ref{fig:ml classifier} illustrates, MDT is initially trained with a Feed Forward Network (FFN) classifier. The inputs (MTS) are fed into the MDT feature extractors to obtain latent states. The latent state is further classified by a ML model.

\section{Experimental Results}
\label{sec:experiments}
This section provides an overview of datasets and their numerical results. The datasets utilized in this article include SCVIC-TS-2022 created by TS-NFM and two other widely-used MTS datasets (i.e., ECG and Wafer) in order to compare MDT to other approaches and demonstrate its generalizability.

\subsection{Datasets}

\subsubsection{SCVIC-TS-2022}

To apply TS-NFM, a network intrusion dataset must have original raw network packets rather than extracted features and complete labeling information; consequently, this paper uses the CIC-IDS-2017 dataset. The raw network packets (PCAP format) are fed into the proposed TS-NFM with a time window of two minutes (same to the extracted features from CIC-IDS-2017), resulting in the SCVIC-TS-2022 dataset. MTS's maximum length $L$ is 511,681 due to the network's fast speed. The number of features/dimensions $d$ is thirteen, which includes the direction of a packet, IAT, size in bytes, and ten TCP flags. 


\subsubsection{ECG}
In order to show the generalizability of the proposed method, we apply the MDT to two more datasets in other domains that have been extensively studied in the early detection research.

The ECG database is made of multivariate time series, each of which represents the sequence of measurements taken by a single electrode during a single heartbeat. Each heartbeat is classified as normal or pathological. All aberrant heartbeats are indicative of a heart condition called a supraventricular premature beat. The ECG dataset contains 200 samples: 133  normal and 67 abnormal. ECG MTS consist of two dimensions/features, with a maximum length of 152.

\subsubsection{Wafer}
The wafer dataset is a collection of MTS including the sequence of measurements taken by a single vacuum-chamber sensor during the each process used to create semiconductor microelectronics. Each wafer is classified as normal or abnormal. The irregular wafers are indicative of a variety of issues that frequently occur during semiconductor fabrication. There are 1194 instances in the Wafer dataset, 1067 of which are normal and 127 of which are abnormal. Wafer MTS have a maximum length of 198 and six dimensions/features.

\subsection{Experimental Results}

This section discusses the detailed results of MDT on the SCVIC-TS-2022 dataset.
To compare with others' work, accuracy, detection rate, and macro (non-weighted) F1 score are used. F1 score of each class can be defined as $F_1 = 2 \frac{Precision*Recall}{Precision+Recall}$. 

\begin{figure}
    \centering
    \includegraphics[width=\linewidth]{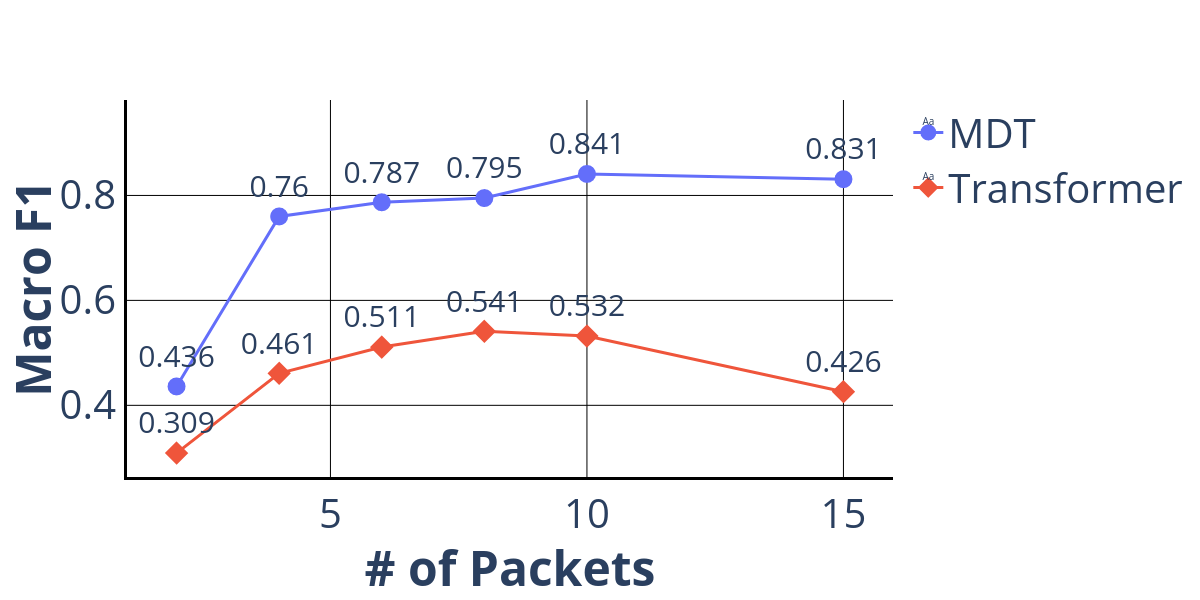}
        \caption{ 
        F1 score Under Varying Number of Packets
        }
    \label{fig:tokens}
\end{figure}

To better understand the influence of earliness and the efficacy of FFT, this study conducts an experiment regarding with earliness/duration-based earliness. Given that the $L$ of SCVIC-TS-2022 is 511,681, the earliness is exceedingly small, ranging from 4e-6 to 2.3e-4; thus, for better depiction, the number of packets is used as the x-axis rather than the earliness. 
Fig.~\ref{fig:tokens} shows that the F1 score is less than 70\% when the first two packets are employed. The F1 score increases as the number of packets increases. When the number of packets exceeds 10, the F1 score can exceed 80\%. 

\begin{figure}
    \centering
    \includegraphics[width=\linewidth]{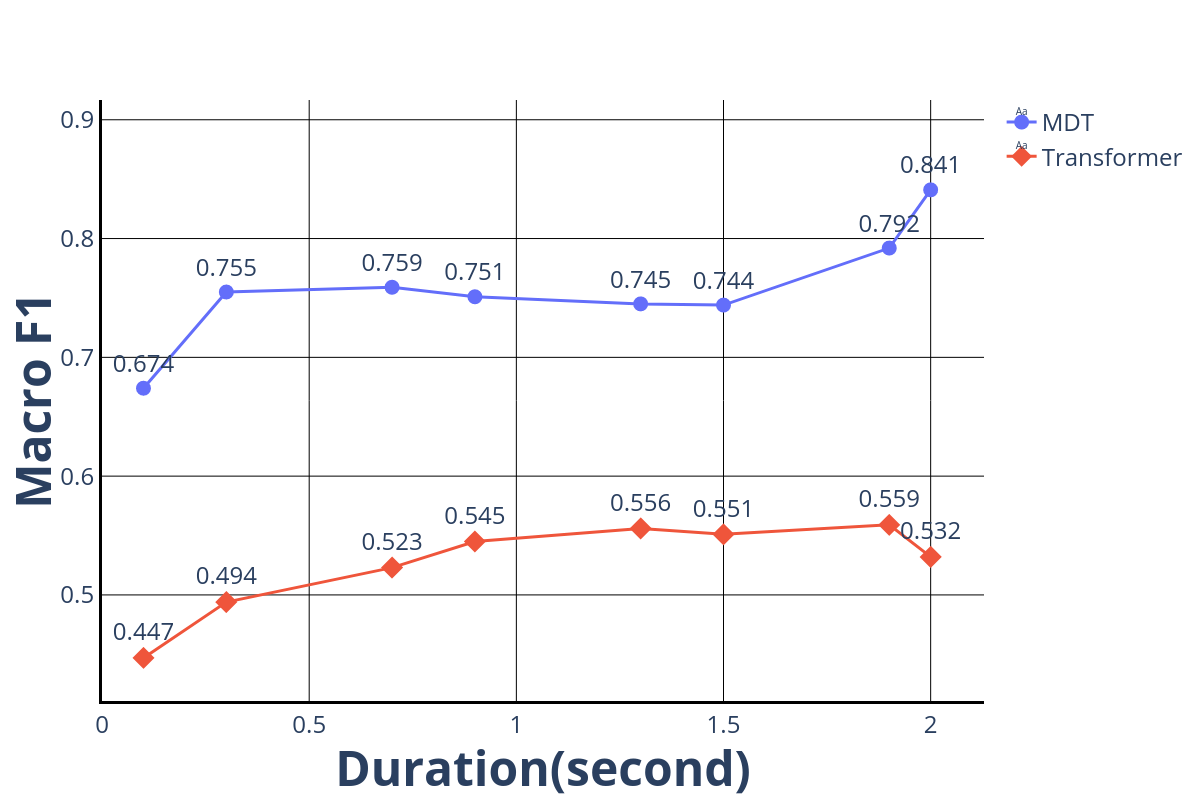}
    \caption{
    F1 score Under Varying Durations 
    }
    
    \label{fig:duration}
\end{figure}


This study compares the performance of Transformer and MDT, and the results demonstrate that Transformer is outperformed by MDT by 12\% to 40.8\% in terms of macro F1 score because it is unable to extract usable features from a small number of packets, FFN struggles to handle unbalanced classes, and oversampling produces significant overfitting. MDT performs significantly better than Transformer as the number of packets increases.
Additionally, Fig.~\ref{fig:duration} illustrates the change in 
F1 score associated with duration-based earliness. When the first 0.1s of packets are used, MDT can get a 67.4\% F1 score. MDT's performance is enhanced when a longer duration is employed. Both Fig.~\ref{fig:tokens} and Fig.~\ref{fig:duration} demonstrate that MDT can consistently outperform Transformer without FFT. At 0.016 duration-based earliness, the maximum improvement of 30.9\% is achieved. As duration-based earliness impacts classification results, it is important to apply it as another evaluation metric. During the experiments, we also notice that FFT can accelerate the convergence of MDT.

\begin{table}[]
\caption{Comparison of the Proposed Method and Related Studies in terms of Detection Performance, Earliness (E) and Duration-based Earliness (DE), Accuracy (ACC); and Detection Rate (DR)}
\label{tab:cicids2017 comparison}
\resizebox{\linewidth}{!}{%
\begin{tabular}{|l|l|l|l|l|l|}
\hline
Method                           & Feature Type & E & DE & ACC (\%) & DR   (\%) \\ \hline
GA-based Adaptive Method \cite{resende_adaptive_2018}          & Tabular      & 1 & 1  & -             & 92.85                 \\ \hline
Ensemble SVM    \cite{marir_distributed_2018}                  & Tabular      & 1 & 1  & -             & 94.94                 \\ \hline
IFSE-AD \cite{wang_using_2018}                          & Tabular      & 1 & 1  & 97.3          & -                     \\ \hline
SU-IDS \cite{min_su-ids_2018}                           & Tabular      & 1 & 1  & 71.02         & -                     \\ \hline
FC-Net \cite{xu_method_2020}                           & Image-like   & 1 & 1  & 94.64         & 99.62                 \\ \hline
\textbf{MDT (Ours)} & \textbf{MTS} & \textbf{2e-5} & \textbf{1.6e-2} & \textbf{99.7} & \textbf{99.7} \\ \hline
\end{tabular}%
}
\end{table}

Although the focus of this paper differs from existing efforts, SCVIC-TS-2022 uses the same network packets as the original CICIDS2017 dataset and has the same flow timeout (2 minutes). Thus, their samples/flows are comparable. Table \ref{tab:cicids2017 comparison} compares MDT to other studies in terms of earliness, duration-based earliness, and performance (i.e., accuracy and detection rate). Tabular features denote the original CICIDS2017's statistical features in CSV format. The use of network packet headers as inputs results in image-like features. In comparison to other approaches, our proposed method increases earliness by 50,000 times and duration-based earliness by 60 times; additionally, our method improves the baseline performance. MDT achieves a higher accuracy of 99.7\% than other baselines and a (weighted-average) detection rate of 99.7\%. Since SCVIC-TS-2022 does not provide any additional feature types compared with CICIDS2017, the results demonstrate that MDT is capable of extracting more valuable information using a very short duration and a limited amount of network packets.

\begin{table}
\centering
\caption{Earliness and F1 score for each method and dataset; E: earliness; DE: Duration-based Earliness.}
\label{tab:results}
\resizebox{\linewidth}{!}{%
\begin{tabular}{|l|l|l|l|l|l|l|l|}
\hline
\diagbox{Method}{Datasets}             & \multicolumn{3}{c|}{SCVIC-TS-2022}               & \multicolumn{2}{c|}{ECG} & \multicolumn{2}{c|}{Wafer}    \\ \hline
           & E & DE & F1 score & E     & F1 score       & E     & F1 score \\ \hline
MSD \cite{ghalwash_early}        & -         & -                  & -       & 0.08          & 0.59          & -             & -       \\ \hline
REACT \cite{lin_reliable_2015}     & -         & -                  & -       & \textbf{0.06} & 0.77          & \textbf{0.23} & 0.92    \\ \hline
1NN-full \cite{ding_querying_2008}  & -         & -                  & -       & 1             & 0.79          & 1             & 0.87    \\ \hline
MDDNN \cite{huang_multivariate_2018}      & -         & -                  & -       & \textbf{0.06} & 0.81          & \textbf{0.23} & 0.91    \\ \hline
ETSCM \cite{hsu_multivariate_2019}      & -         & -                  & -       & 0.13          & 0.89 & 0.5           & 0.93    \\ \hline
\textbf{MDT (Ours)} & \textbf{2e-5} & \textbf{0.016} & \textbf{0.84} & \textbf{0.06}   & 0.84   & \textbf{0.23} & \textbf{0.98} \\ \hline
\textbf{MDT (Best F1 score)} & \textbf{2e-5} & \textbf{0.016} & \textbf{0.84} & 0.4   & \textbf{0.94}   & \textbf{0.23} & \textbf{0.98} \\ \hline
Transformer & 2e-5    & 0.016              & 0.53    & \textbf{0.06} & 0.59          & \textbf{0.23} & 0.94    \\ \hline
\end{tabular}%
}
\end{table}

Table \ref{tab:results} compares our proposed MDT to five other state-of-the-art approaches in terms of earliness and F1 score. ETSCM is likewise inspired by the MDDNN; however, rather than using the earliness of earlier work, ETSCM chooses earliness based on its best F1 score. Thus, it is compared to MDT's best F1 score. REACT and MDDNN have the best earliness for ECG datasets. MDT achieves an 84\% F1 score, which is 3\% higher than MDDNN when the same earliness (0.06) is used. In comparison to ETSCM, it achieves the greatest F1 score of 89\% at 0.13 earliness, while MDT obtains the highest F1 score of 94\% (5\% better than ETSCM) at 0.4 earliness. Due to the fact that the ECG dataset contains only two features, it is difficult for Transformer to analyze, providing 59\% F1 score; as a result, MDT improves Transformer by 25\% on the ECG dataset. 
Previous work achieved 92\% F1 score within 0.23 earliness for the Wafer dataset, whereas the proposed MDT achieves 98\% F1 score under the same earliness with an improvement of 6\%. Even when compared to the best performance of ETSCM, we are able to achieve a 5\% increase in F1 score with better earliness. 
The results indicate that MDT with FFT 
improves the Transformer's F1 score by 4\% under the Wafer dataset.

\section{Conclusion}
\label{sec:conclusion}

NIDSs have been extensively investigated but traditional approaches take a relatively long time to form flows and accumulate distinguishable features (e.g., 2 minutes for CICIDS2017). 
This work introduces early detection into the NIDS field, allowing for the detection of malicious traffic before it reaches target systems in its whole. This paper introduces a novel feature extractor, TS-NFM, that describes network flow as MTS with explainable features, and generates the SCVIC-TS-2022 dataset; additionally, a new early detection model, Multi-Domain Transformer (MDT), is proposed. When compared to the conventional approach, our proposed method improves earliness by 50,000 times and duration-based earliness by 60 times. This study reveals that when the first ten packets in 2s are used for detection, the proposed method is capable of achieving satisfactory detection performance (higher than 84\% F1 score). To demonstrate MDT's generalizability further, two additional datasets have been analyzed. MDT is capable of improving the F1 score of state-of-the-art algorithms by 5\% and 6\%, respectively, under the ECG and Wafer datasets.
Our ongoing study involves evaluation of the results under multiple network intrusion datasets.

\section*{Acknowledgment}
This work was supported in part by the Natural Sciences and Engineering Research Council of Canada (NSERC) under Grant RGPIN/2017-04032.

\bibliographystyle{IEEEtran}

\end{document}